\documentclass[aps,prl,twocolumn,letterpaper,showpacs]{revtex4-1}
\usepackage{graphicx,amsmath,amssymb,mathrsfs,amsthm}

\usepackage[usenames,dvipsnames]{color}

\newcommand{ \ket }			[1] { \left|{#1}\right> }

\begin{document}
\pagenumbering{arabic}
\title{Failure of protection of Majorana based qubits against decoherence}
\author{Jan Carl Budich$^1$}
\author{Stefan Walter$^1$}
\author{Bj\"orn Trauzettel$^1$}

\affiliation{$^1$Institute for Theoretical Physics and Astrophysics,
University of W$\ddot{u}$rzburg, 97074 W$\ddot{u}$rzburg, Germany}

\date\today

\begin{abstract}
Qubit realizations based on Majorana bound states have been considered promising candidates for quantum information processing which is inherently inert to decoherence. We put the underlying general arguments leading to this conjecture to the test from an open quantum system perspective. It turns out that, from a fundamental point of view, the Majorana qubit is as susceptible to decoherence as any local paradigm of a qubit.
\end{abstract}
\maketitle

Recently, proposals for topological quantum computing (TQC) with qubits based on Majorana bound states (MBS) as realized in one dimensional (1D) topological superconductors (TSC) have attracted a lot of interest
\cite{Kitaev:2001p131,Oreg:2010p177002,Alicea:2010p125318,Lutchyn:2010p077001,Sau:2010p254,Alicea:2011p260,Hassler:2010p259,Hassler:2011p299,Leijnse:2011p284,Flensberg:2011p279,MartinMorpurgo2011}.
These 1D TSC have a bulk superconducting gap and support a single subgap fermionic state $f$~which is formed by a single delocalized pair of MBS: one MBS at the left end ($\gamma_{L}$) and one MBS at the right end ($\gamma_{R}$) of the 1D TSC.  The two Majoranas combine to this one ordinary Dirac subgap fermion: $\gamma_{R} = f^{\dag} + f$ and $\gamma_{L} = -i \left( f^{\dag} - f \right)$. This class of systems has originally been proposed and topologically classified by Kitaev \cite{Kitaev:2001p131,Kitaev:2009p296}. The protected existence of the single pair of MBS is due to a nontrivial value of the $\mathbb Z_2$~invariant classifying a 1D bandstructure in the presence of particle hole symmetry \cite{Kitaev:2009p296}. The qubit formed by the two occupation number eigenstates of the single subgap fermion $f$~has been recently proposed as a candidate for TQC \cite{Alicea:2011p260}. In this work, we show that while the existence of a single pair of MBS in a 1D TSC is protected, the coherence of the associated qubit is as vulnerable as that of an ordinary local fermionic subgap bound state. We first review the two general remarks in Ref. \cite{Kitaev:2001p131} supporting the protection of this qubit against any local perturbation, a crucial prerequisite for TQC: 

{\it Remark} $(i)$: The qubit is delocalized into the two MBS $\gamma_L,~\gamma_R$~which are spatially separated by the system length $L$. Since the overlap of the bound state wave function decays exponentially with the system length, direct coupling between the two MBS can be suppressed to exponential accuracy.

{\it Remark} $(ii)$: Fermion parity, i.e. particle number conservation modulo 2, is a good quantum number in the superconducting system. Thus, any perturbation containing a single Majorana operator  is forbidden as its action would change the fermion parity of the TSC.

Now, we want to investigate whether these key observations for a closed, noninteracting TSC still hold in an open quantum system scenario which is the only realistic approach to describe an actual experimental setup for quantum information processing.

{\it Discussion of Remark} $(i)$. For a system consisting of two entangled spatially separated subsystems, the existence of states where information about the composite system can be inferred by locally coupling to one subsystem due to the mutual information of the entangled constituents, has been known for many decades. Furthermore, ground state entanglement and topological order are in close correspondence \cite{Levin:2006p308, Kitaev:2006p318, Chen:2011p292, Fidkowski:2011p307}. Several recent proposals \cite{Semenoff:2007p1479,Tewari:2008p301,Fu:2010p258,Bose:2011p322} related to teleportation between the two MBS could demonstrate how a local operation on one side of the system changes the system state nonlocally even in the limit $L\rightarrow \infty$ \cite{Semenoff:2007p1479,Bose:2011p322}, where the direct overlap and with that the direct coupling between the end states vanishes. In this sense a vanishing direct coupling between the two MBS does not imply that the information of the qubit is split into two independent halves.

{\it Discussion of Remark} $(ii)$. The susceptibility to decoherence of any candidate system has to be investigated from an open quantum system point of view since decoherence is the elusion of coherence to a larger Hilbert space of the combined qubit-environment system. Considering only the isolated qubit system the absence of decoherence would be a trivial corollary from the unitarity of its time evolution. From this point of view the practical relevance of {\it Remark} $(ii)$ is not very convincing as it only pertains to the TSC representing the qubit as an isolated system. In presence of an environment which is particle number conserving or at least fermion parity conserving, the only constraint on the dynamics of the total system is the conservation of the total fermion parity. Operations like particle tunneling conserve the total fermion parity but change the parity of each subsystem, the TSC and the environment. Hence, it is not surprising that several proposals
\cite{Bolech:2007p237002,Tewari:2008p301,Law:2009p237001,Flensberg:2010p180516,Shivamoggi:2010p309,Fu:2010p258,Golub:2011p283,Leijnse:2011p280,Zazunov:2011p282,Liu:2011p281,Stanescu:2011p264,Walter:2011p295,Bose:2011p322}
use such couplings to probe the properties of MBS by tunneling based transport experiments. In the limit of a large superconducting gap the only low energy degrees of freedom are the two degenerate ground states $\lvert 0\rangle,~\lvert 1\rangle= f^\dag \lvert 0\rangle$~of the wire forming the qubit. Tunneling between an electron from the environment and this subgap fermion will thus inevitably flip the information stored in the parity qubit, i.e. lead to $\sigma_x$~errors. Unless any fundamental reason beyond the parity argument by Kitaev can be found that such couplings are weaker than sources of decoherence in any alternative realization of a qubit, there is no topological protection against decoherence in the MBS paradigm of a qubit to speak of. Considering these rather general arguments it is again not surprising, that recently the vulnerability of the MBS qubit to several concrete mechanisms of decoherence has been demonstrated \cite{Goldstein:2011p278}.\\
We now illustrate the fragility of the parity qubit with the help of two minimal toy models for imperfections which will be present in any realistic experimental setup for topological quantum computing.

\begin{figure}[ht]
	\center{\includegraphics[width=0.9\columnwidth]{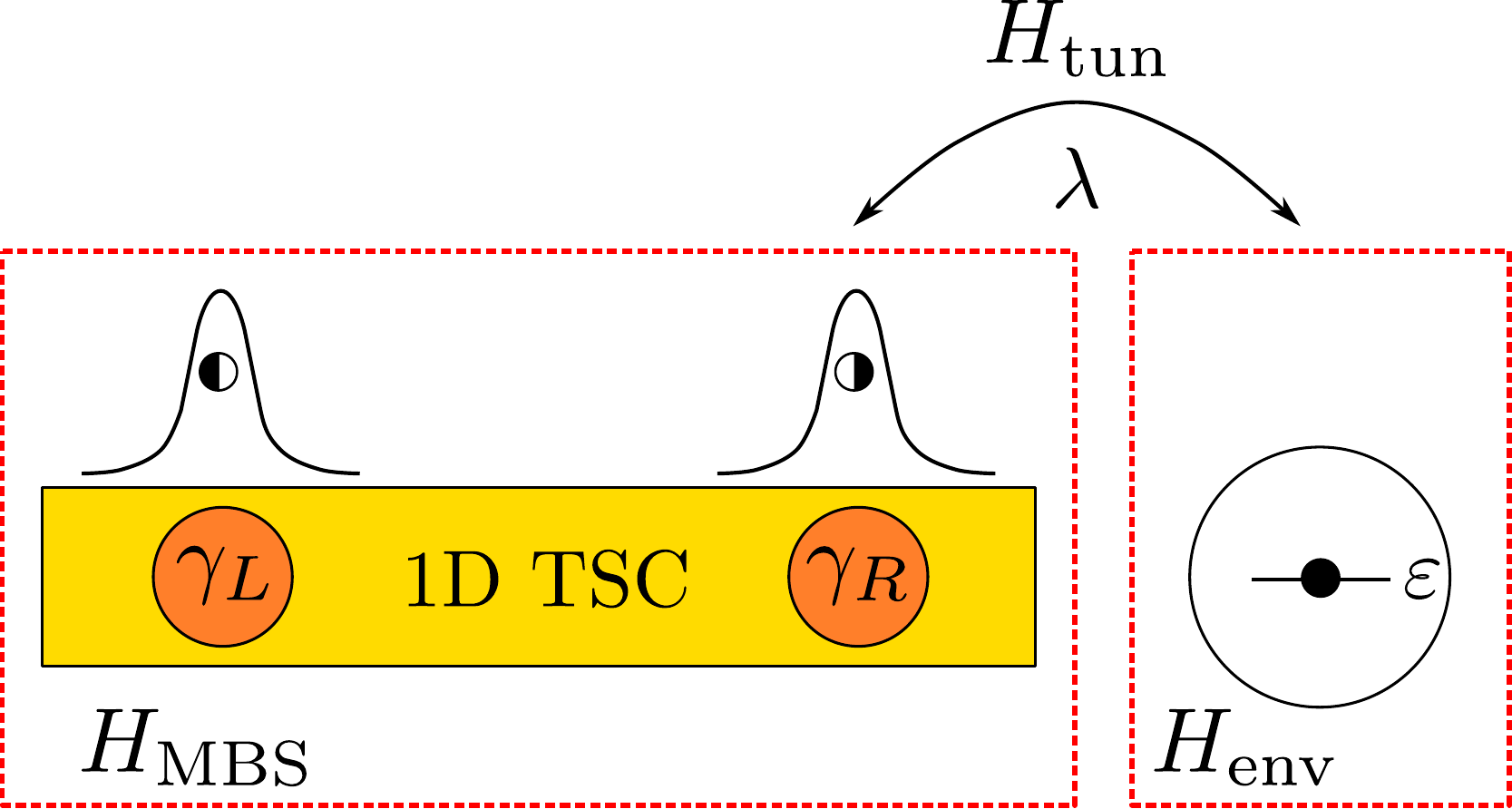}}
	\caption{\label{fig:schematic} (Color online) TSC tunnel coupled to its environment here represented by a single level dot as toy model for a surface adatom.}
\end{figure}

Both toy models are described by a similar Hamiltonian $H = H_{\rm{env}} + H_{\rm{MBS}} + H_{\rm{tun}}$, with $H_{\rm{env}}$ being the Hamiltonian of the environment and $H_{\rm{MBS}} = i \xi \gamma_{L} \gamma_{R}/2 = \xi f^{\dag}f $
describing the overlap between MBS at the left and right edge of the 1D TSC.
$H_{\rm{tun}}$ is a tunnel Hamiltonian coupling the MBS and the environment and will be specified for each toy model. Note that no orthogonality catastrophe forces $H_{\rm{tun}}$ to have a vanishing tunneling matrix element in the thermodynamic limit. This can be seen by an explicit derivation of the tunneling Hamiltonian as, for instance, done in Ref.~\cite{Flensberg:2010p180516}. The toy models resemble typical physical situations which are present in the 1D TSC wire (e.g. adatoms or trapped charges nearby the MBS) or are
induced from the outside to the wire (biased gates near the MBS to manipulate the MBS as e.g. in Ref.~\cite{Alicea:2011p260}) . \\
% DOT
The first toy model schematically shown in Fig. \ref{fig:schematic} is a single level quantum dot tunnel coupled to the parity qubit, see also Ref.~\cite{Leijnse:2011p280}, here without the spin degree of freedom for simplicity. Such a two level system describes e.g. a minimal model for trapped charges
nearby the MBS in the wire which is allowed by symmetry. With $H_{\rm{env}} = \varepsilon d^{\dag} d$ and the dot only coupling to $\gamma_{R}$ via the following tunnel Hamiltonian
\begin{align}\label{eqn:M1}
	H_{\rm{tun}} = \lambda \left[ d^{\dag} - d \right] \gamma_{R},~\lambda\in \mathbb R .
\end{align}
The low energy Hamiltonian $H$ can be conveniently written as a matrix choosing the basis
$\{ \ket{00},\ket{10},\ket{01},\ket{11} \}$
\begin{align}\label{eqn:M2}
	H = \left(\begin{array}{cccc} 0 & 0 & 0 & \lambda \\ 0 & \varepsilon &\lambda & 0 \\ 0 & \lambda & \xi & 0 \\\lambda & 0 & 0 & \varepsilon+\xi \end{array}\right) \, ,
\end{align}
with $\ket{n_{\rm{dot}} \, n_{f}}$ and $n_{{\rm{dot}},f} \in \{ 0,1 \}$ being the occupation number of the single dot level and the MBS qubit, respectively. In order to investigate decoherence of the parity qubit, we study the time evolution of the
reduced density matrix for the MBS qubit $\rho_{f}(t) = \rm{Tr}_{\rm{dot}}[ \rho(t) ]$ which can readily be solved exactly. The time evolution of the density matrix of the full system reads $\rho(t) = e^{-iHt} \rho(0) e^{iHt}$, where we have set $\hbar=1$. As an
example, we consider the time evolution of the parity qubit's occupation number $n_{f}$ for an initially occupied dot and an empty subgap fermion state for $\varepsilon=\xi=0$. The MBS qubit performs Rabi oscillations of full amplitude and is thus totally unstable on the time scale given by the coupling strength $\lambda$. The revivals of the initial state are of course due to the finite number of environmental degrees of freedom. However, since the precise number of imperfections, coupling parameters etc. are not experimentally accessible, the reduced state of the qubit to be finally read out will become totally unpredictable due to this kind of environmental coupling. 

Furthermore, the coupling between a trapped charge and the TSC can change during a braiding operation, and thus lead to unwanted errors. Look, for instance, at the operations proposed in Ref.~\cite{Alicea:2011p260}. A trapped charge might be close to one of the arms of the wire network. If the MBS is located (during an operational step) in that arm it is coupled to the trapped charge via electron tunneling and if not it does not feel its presence. Hence, during a braiding operation, the tunnel coupling could be unintentionally turned on and off. It needs to be analyzed how this kind of error may affect the success of braiding operations.

The second model we investigate consists of a 1D TSC tunnel coupled on one end to a metallic lead which might be realized by a gate or a tip used to implement operations on the qubit. We assume a very long 1D TSC and therefore concentrate on the MBS at the right edge. The lead Hamiltonian $H_{\rm{env}} = \sum_{k} \varepsilon(k) \psi_{k,R}^{\dag} \psi_{k,R}$ and the coupling is given by
\begin{align}\label{eqn:M3}
	H_{\rm{tun}} &= \lambda_{R} \left[ \psi_{R}^{\dag}(x=0) - \psi_{R}(x=0) \right] \gamma_{R} ,~ \lambda_R\in \mathbb R ,
\end{align}
and we assume a linear dispersion $\varepsilon(k)$~for simplicity.
In the following, we study the spectral function of the MBS at the right edge $A(\omega) = - 2 \rm{Im}\left[ G^{R}_{\gamma_{R}\gamma_{R}}(\omega) \right]$, where the retarded Green's function of the MBS
is calculated solving the full non-equilibrium Dyson equation on the Keldysh contour. For the sake of brevity, details of the calculation which is straight forward are omitted here. After Fourier transforming back to the time domain, we obtain
\begin{align}\label{eqn:M4}
	A(t>0) = e^{-4\pi \rho_{0} \lambda_{R}^{2} t} \, ,
\end{align}
where $\rho_{0}$ is the constant density of states in the metallic lead.
The lifetime of the Majorana bound state is thus determined by the tunnel coupling and the density of states offered for tunneling by the environment, similarly to any local qubit exposed to tunnel coupling. In particular, an ordinary local fermionic subgap bound state would behave very similar when tunnel coupled to its environment. Of course, the spectral weight of our MBS based state $f$~is delocalized over the two ends of the TSC, but this would only lead to a reduction of the tunnel coupling by a factor of $1/\sqrt{2}$~as compared to a local bound state.\\
Although the Majorana qubit is defined as a nonlocal object, local coupling to a MBS via a tunnel Hamiltonian as in Eqs.~(\ref{eqn:M1}) and (\ref{eqn:M3}) is extensively studied in the literature \cite{Bolech:2007p237002,Tewari:2008p301,Law:2009p237001,Flensberg:2010p180516,Shivamoggi:2010p309,Fu:2010p258,Golub:2011p283,Leijnse:2011p280,Zazunov:2011p282,Liu:2011p281,Stanescu:2011p264,Walter:2011p295,Bose:2011p322}, particularly as a way to detect the MBS. We would like to point out that such a coupling already contradicts the fundamental conjecture $(ii)$~which is crucial for TQC in MBS based systems. While the presence of subgap MBS is topologically protected by particle hole symmetry \cite{Kitaev:2001p131,Kitaev:2009p296}, TQC tasks with MBS
as proposed, for instance, in \cite{Hassler:2010p259,Alicea:2011p260,Flensberg:2011p279,Leijnse:2011p284} are not protected against decoherence by any fundamental symmetry in particular not by a topological one.

\begin{figure}[ht]
	\center{\includegraphics[width=0.9\columnwidth]{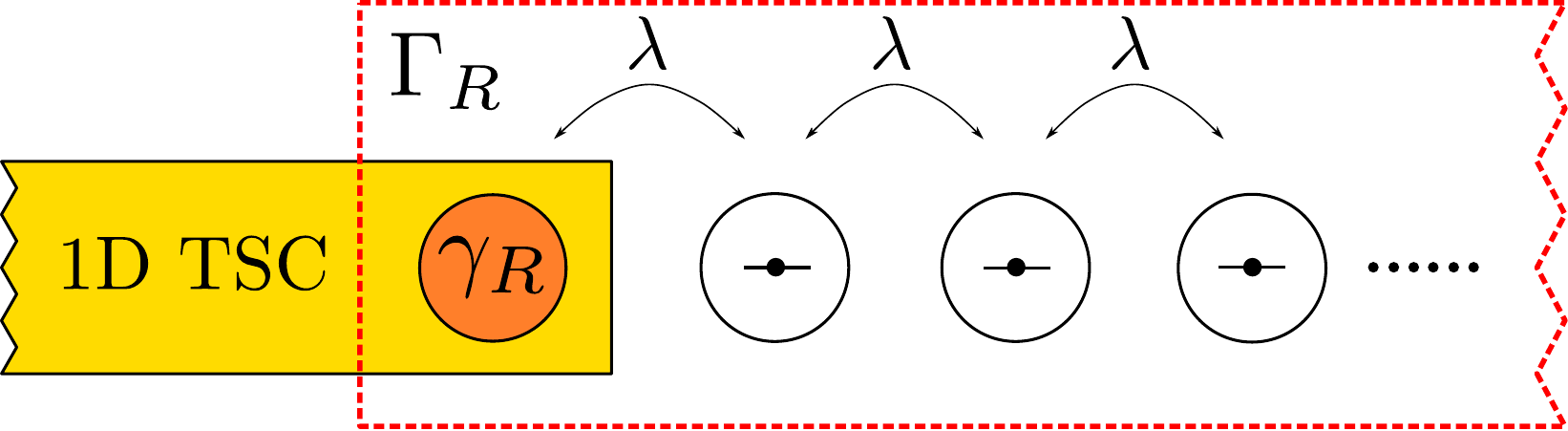}}
	\caption{\label{fig:largequbit} (Color online) Right end of a 1D TSC coupled to environmental degrees of freedom. To obtain a closed coherent qubit, in general, an uncontrollable number of environmental states would have to be included into the qubit.}
\end{figure}

An interesting idea to practically improve the stability of the MBS qubit has been presented in Ref. \cite{Akhmerov:2010p303} where a qubit consisting of the total fermion parity of the MBS pair $\gamma_L,~\gamma_R$~and of some additional fermionic states bound to the Majorana vortices is considered by defining:
\begin{align}
\Gamma_i = \gamma_i\prod_{j}\left(1-2\psi_{i_j}^\dag\psi_{i_j}\right),
\end{align}
with $\gamma_i$~the MBS operator and $\left\{\psi_{i_j}\right\}_j$~the environmental operators coupling to $\gamma_i$. However, generally speaking, including the bath into the system to trivially obtain coherence on the total composite system is definitely not an experimentally viable approach, not even in principle (see Fig. \ref{fig:largequbit} for a schematic).
Furthermore, we point out that an improved stability arising from such a procedure, if possible for some special cases of a controllable coupling, would protect the parity of a local fermionic bound state qubit as well. Thus it cannot be considered as a topological protection of a delocalized MBS based qubit.\\

Up to now we have concentrated on 1D TSC as a MBS qubit system. As already mentioned in Ref. \cite{Goldstein:2011p278}, similar arguments apply for any system where quantum information is stored in MBS.
In fact, most of the experimentally relevant proposals for topological quantum computing with nonabelian anyons are working with Ising anyons which are based on MBS \cite{Nayak:2008p51}.
The general concept of TQC relies on the following crucial observation. If the low energy theory of the physical system representing the quantum computer is a topological field theory (TFT), there is an inherent robustness of the system against any local perturbation. In the framework of TFT this observation is trivial since there is no physical length scale in the system on which any local correlation could occur. However, this low energy theory for the candidate system including generic unavoidable imperfections, e.g. adatoms on its surface etc., is often times not derived from first principles. Therefore, robustness against decoherence from an open quantum system point of view requires the validity of the following statement. The system including, for instance, also the STM tips proposed to create anyonic quasiparticles \cite{Nayak:2008p51} and generic imperfections present in any experimental setup must be represented by a TFT with nonabelian anyons being the only low energy degrees of freedom separated from all other excitations by a sufficiently large gap. Otherwise the protection though manifest in the TFT describing the ideal system is of no practical relevance as uncontrollable low energy degrees of freedom might be present in the coupled system.

To sum up, as far as MBS based qubits are concerned we gave two general reasons why the protection against decoherence will fail for quite mundane coupling mechanisms. In particular there is no fundamental difference in the stability of the fermion parity for a MBS pair and a local fermionic bound state which is separated from bulk excitations by a superconducting gap. These results have been established by critically revisiting the crucial ingredients for TQC in 1D TSC from a general open quantum system perspective and have then been illustrated with the help of two minimal toy models. The topological protection in a 1D TSC thus pertains to the presence of the single pair of MBS and not to the coherence of the associated qubit. Furthermore, since braiding operations are in a closed system independent of the local details of the path traversed by the quasiparticles, the precision of these operations is not sensitive to the mechanical fine tuning of the control ports of the setup. This feature is of course not related to the coherence properties of a candidate system for TQC. The usefulness of MBS based quantum computers will thus be decided by practical aspects of material science rather than by fundamental arguments related to nonlocal storing of information: Can particle exchange be suppressed much more efficiently then other mechanisms leading to decoherence of say the spin of a trapped ion or a quantum dot or the phase of a flux qubit? Comparing different approaches on this rather applied level a strong argument supporting many alternative approaches to quantum computing , see e.g. Refs \cite{Cirac:1995p326,Loss:1998p325,Wallraff2004}, is that their basic constituents are readily experimentally accessible.\\

We thank Daniel Loss, Naoto Nagaosa, and Pauli Virtanen for interesting discussions and constructive comments. Financial support by the DFG is gratefully acknowledged.
% \bibliographystyle{apsrev4-1}
% \bibliography{telemaj}
%
\end{document}